# A valley valve and electron beam splitter

Jing Li[1], Rui-Xing Zhang[1], Zhenxi Yin[1], Jianxiao Zhang[1], Kenji Watanabe[2], Takashi Taniguchi[2], Chaoxing Liu[1], Jun Zhu[1,3]

**Affiliations:**

[1]Department of Physics, The Pennsylvania State University, University Park, Pennsylvania 16802, USA.

[2]National Institute for Material Science, 1-1 Namiki, Tsukuba 305-0044, Japan.

[3]Center for 2-Dimensional and Layered Materials, The Pennsylvania State University, University Park, Pennsylvania 16802, USA.

*Correspondence to: jzhu@phys.psu.edu (J. Zhu)

**Abstract:**

Developing alternative paradigms of electronics beyond silicon technology requires the exploration of fundamentally new physical mechanisms, such as the valley specific phenomena in hexagonal two-dimensional materials. We realize ballistic valley Hall kink states in bilayer graphene and demonstrate gate-controlled current transmission in a four-kink router device. The operations of a waveguide, a valve and a tunable electron beam splitter are demonstrated. The valley valve exploits the valley-momentum locking of the kink states and reaches an on/off ratio of 8 at zero magnetic field. A magnetic field enables a full-ranged tunable coherent beam splitter. These results pave the path to building a scalable, coherent quantum transportation network based on the kink states.

**Main Text:**

The advent of two-dimensional layered materials such as graphene and transition metal dichalcogenides has inspired the concept of devices, that exploit the valley degrees of freedom in materials with hexagonal symmetry (*1-4*). Experiments have shown that a valley polarization can be created by current (*5-7*) or optical excitation (*8*). However the realization of valleytronic devices remains challenging. In bilayer graphene, a perpendicular electric field applied through a pair of top and bottom gates breaks the symmetry of the two constituent layers and opens a gap $\Delta$ in its band structure (*9, 10*). This gap can be inverted by switching the direction of the applied electric field; if two electric fields of opposite sign are applied on two neighboring regions in the sample, metallic, helical, quantum valley Hall kink states (kink states for short) emerge along the zero gap line (*11*). Topological in origin, the kink states are chiral in each valley and have opposite chiralities, i.e. group velocities, in valley K and K′ (-K) (Fig. 1A). They are expected to be immune from backscattering in the absence of valley-mixing scattering events, and thus capable of carrying current ballistically over long distances without dissipation (*11-18*).

The intrinsic properties of the kink states enable several in situ transmission control mechanisms. Figure 1A illustrates the generation of kink states in bilayer graphene through asymmetric gapping (*11, 12*). The shown chiralities correspond to the (- +) gating configuration. A (+ -) configuration simultaneously flips the chirality of the kink states in both valleys. Symmetric gapping configurations (+ +) or (- -) do not produce kink states and serve as controls in our experiment (Fig. S3). The existence of two helicities produced by the (- +) and (+ -) gating is a unique attribute of the kink states and leads to the proposal of an all-electric valley valve (*12, 14*), the operation of which relies on the valley-momentum locking of the kink states. Different from a classical spin valve (*19*), a valley polarization is not a requirement for the proposed valley valve. Here we show the realization of the valley valve with a transmission on/off ratio of 8 at zero magnetic field and more than 100 at several Teslas.

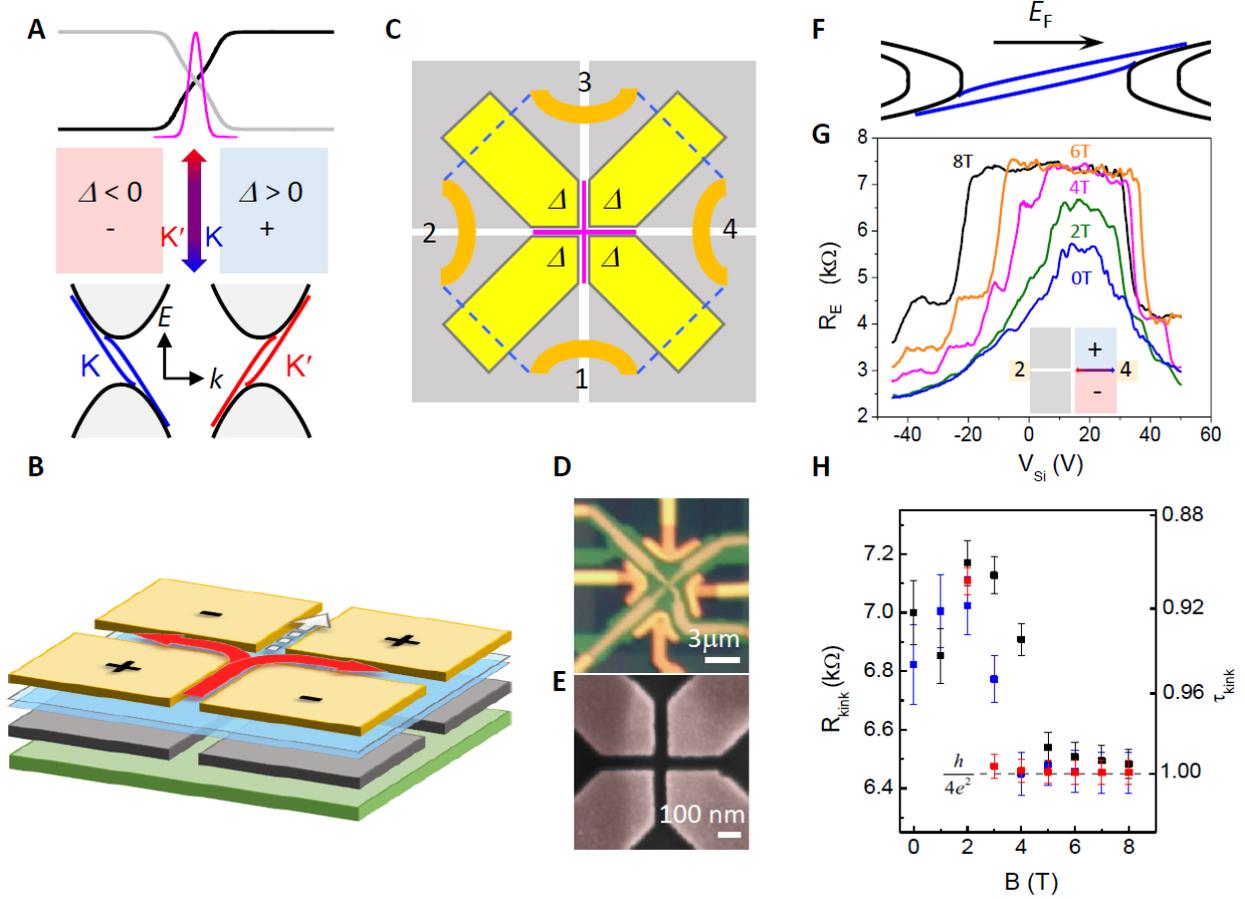

**Fig. 1. A valley router device and ballistic conduction of the kink states.** (**A**)The potential profile and wave function distribution (magenta curve) of the valley-momentum locked kink states in a 70nm-wide junction (*11*). Including spin and layer isospin, there are four chiral modes in each valley. (**B**) and (**C**), Schematics of our quad-split-gated valley router device. The four graphite bottom gates are shown in gray. The four top gates are yellow. The bottom gates are set to ± 3 V with the polarity given in diagrams unless otherwise specified. The top gates are set to place the dual-gated regions at the CNP. 3V corresponds to a bulk gap of $\Delta \sim 86$ meV (see Section 2 of (*27*)). The blue sheet in (**B**) and dashed lines in (**C**) represent the bilayer graphene sheet. The global Si backgate is light green. The four gold arcs are Cr/Au side contacts. The magenta cross in (**C**) represents the four kink channels. Each is 70 nm wide and 300 nm long. The red arrows and white dashed arrow in (**B**) represent the valley valve and beam splitting actions discussed in Fig. 3. (**D**), An optical image of device 1. (**E**) A false colored scanning electron micrograph of the central region taken on another device. (**F**), A band diagram of the junction in magnetic field (*11*). (**G**), $R_E$ ($V_{Si}$) at different magnetic fields as labeled in the graph. $R_E$ is obtained by measuring $R_{24}$ while doping the left quadrants heavily, as illustrated in the inset. (**H**), Resistance of the kink state $R_{kink}$ as a function of the magnetic field in the east (black) and south (blue) channels of device 1 and south (red) channel of device 2. The right axis labels the corresponding transmission coefficient $\tau_{kink}$. $T = 1.6$ K in all our measurements.

Figures 1, B - E show schematics, optical and scanning electron micrographs of the four-terminal valley router device, which consists of four pairs of split top and bottom gates and a global Si backgate (see Fig. S1 for device fabrication). The aligned edges of the eight gates define the four kink channels shown in magenta in Fig. 1C. The dual-gated region (yellow areas in Fig. 1C) in each quadrant is gapped and placed at the charge neutrality point (CNP) (See Fig. S2 for device characterization). We first measure the resistance of each kink channel $R_{kink}$ separately. As an example, Fig. 1G plots the resistance of the east channel $R_E$ as a function of Si backgate $V_{Si}$, which controls the Fermi level $E_F$ in the channel, at fixed magnetic fields $B = 0$ to 8T. $R_E$ exhibits a broad peak at $B = 0$, which evolves into a wide plateau as $B$ increases and saturates at about 7.3 k$\Omega$. This plateaued region corresponds to the gapped regime of the channel, where the kink states reside. We call this kink regime. Its resistance value of 7.3 k$\Omega$ is a sum of the ballistic resistance of the kink states, i.e. $h/4e^2 = 6.5$ k$\Omega$, where $h$ is Planck constant and a contact resistance $R_c \sim 800$ $\Omega$. Additional plateaus outside the kink regime correspond to the sequential addition of 4-fold degenerate quantum Hall edge states in the channel (Fig. 1F). The application of a perpendicular magnet field has little effect on the energy spectrum of the kink states (*11, 20*) and does not generate additional edge states inside the band gap [Fig. S3; (*21, 22*)]. It however turns the conduction and valence bands of the junction into Landau levels (Fig. 1F) as evidenced by the appearance of additional resistance plateaus. The devices studied here are of higher quality than those reported in our previous work (*11*) owing to the adoption of the van der Waals transfer method (*23*). A side effect of this approach, however, is a large width/length ratio of the dual-gated regions (Fig. 1D), which enhances parallel hopping conduction. At small magnetic fields, the associated resistance $R_{para}$ causes $R_E$ to be less than $h/4e^2$ (see curves for $B = 0$ and 2 T in Fig. 1G). We measure $R_{para}$ independently using the symmetric gapping configurations (Figs. S3 and S5) and extract $R_E$ using a two-resistor model $R_E = \frac{(R_{24}-R_C)R_{para}}{R_{para}-(R_{24}-R_C)}$. As $B$ increases, hopping conduction becomes increasingly suppressed and $R_{para}$ grows to hundreds of MΩ. $R_{para}$ becomes inconsequential at 4 T, which leads to the observed saturation of $R_{24} = h/4e^2 + R_c$ in Fig. 1G. We determine $R_c$ by fitting a series of quantized resistance plateaus and have observed robust resistance quantization of the kink states at $h/4e^2$ in different devices (Fig. S4).

Similar measurements and analyses were performed on other channels in the same device and in a second device, the results are shown in Fig. 1H. At $B = 0$, $R_{kink}$ is approximately 7000 Ω for our 300 nm-long channels, which corresponds to a transmission coefficient $\tau_{kink} = h/4e^2 / R_{kink} = 0.92$ and an estimated mean free path of 3.5 μm (*11*). This is on par with the mean free path of the quantum spin Hall edge states (*24, 25*) and affirms the topological protection provided by the valley-momentum locking of the kink states. Because this protection vanishes in the armchair crystallographic orientation, we ensure that neither of the two perpendicular directions of the channels is aligned with either zigzag or armchair orientations in our devices. Indeed, the similar performances of channels perpendicular to one another in our devices (Fig. 1H) support the existence of topological protection for both channels. The presence of a magnet field improves the ballisticity of the kink states, which exhibit a quantized resistance plateau at $B \sim 4$ T. As discussed in our previous work (*11*), we suspect that the backscattering of the kink states is caused by one-dimensional non-chiral states bound in the junction, as well as by charge puddles inside the gap. The application of a magnetic field moves both types of states to higher energy, thus reducing their interactions with the kink states. Additionally, in the 0$^{th}$ Landau level of bilayer graphene, states in K and K′ occupy different graphene layers (*22*). If kink states behave similarly, this could contribute to reduced backscattering as well. A quantitative study can shed more light on this issue.

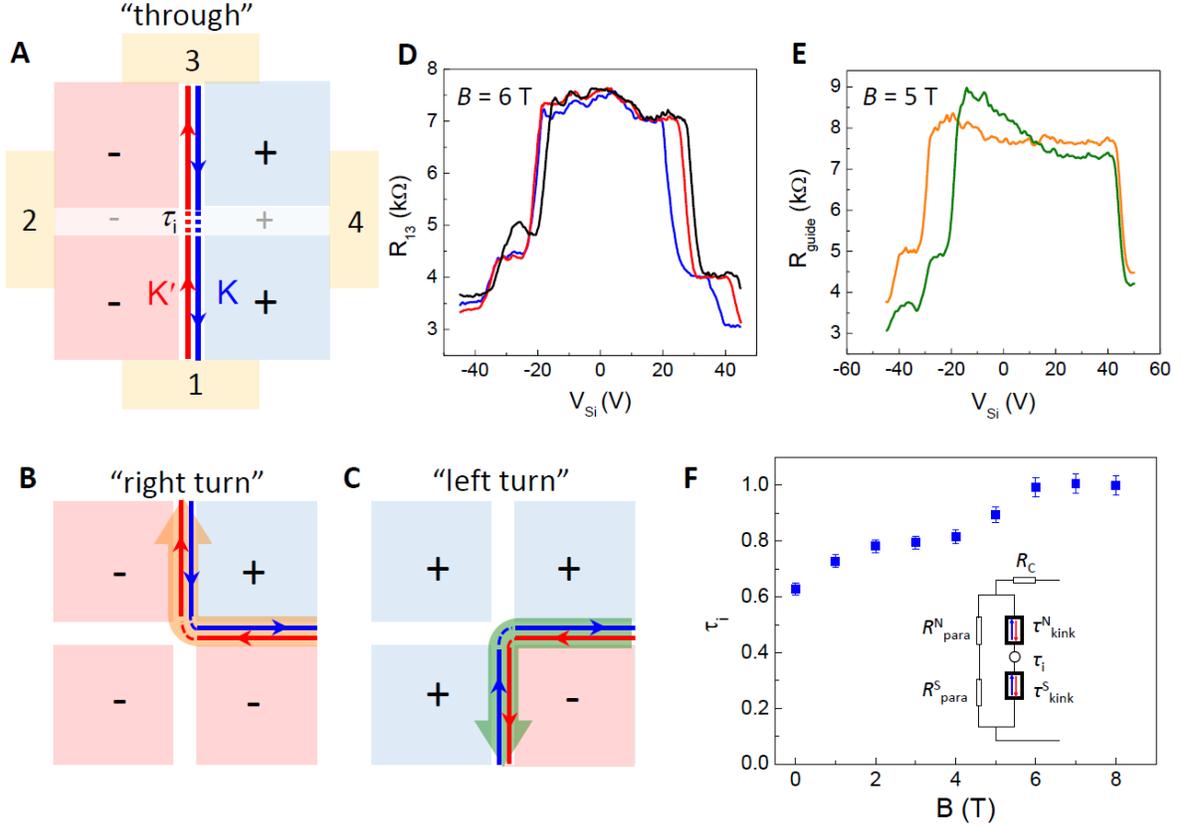

**Fig. 2. Transmission of the kink states in the waveguide mode of the router. (A) – (C),** illustrate the "through", "right turn" and "left turn" configurations of the waveguide respectively. **(D),** Two-terminal resistance $R_{13}$ ($V_{Si}$) corresponding to the resistance of the north channel $R_N$ (blue), the south channel $R_S$ (red) and the "through" configuration (black), respectively. The overlap of all three indicates ballistic transmission through the intersection region in **(A)**, i.e. $\tau_i = 1$. $B = 6$ T. **(E),** $R_{34}$ (orange) and $R_{14}$ (olive) as a function of $V_{Si}$ in the configurations shown in **(B)** and **(C)** respectively. $B = 5$ T. **(F),** The transmission coefficient $\tau_i$ of the intersection region in **(A)** as a function of the magnetic field with the schematic of the two-resistor model shown in the inset.

We now demonstrate the operation of the valley router as a reconfigurable waveguide for the kink states. Figures 2, A to C show three configurations of the waveguide, which we label as "through", "right turn" and "left turn", respectively. In all three, the kink states only exist in two of the four channels and the chirality in each valley is preserved along the paths. Figure 2D plots the measured "through" resistance $R_{13}$, together with the resistance of each individual kink channel $R_N$ and $R_S$ at 6T All three curves overlap in the kink regime, suggesting that the transmission through the intersection region is ballistic. Similar ballistic transmission is also observed in the two 90° bends (Fig. 2E), consistent with the results of numerical simulations (*14, 26*). This four-terminal device thus serves as an in situ reconfigurable electronic waveguide of the kink states. The ability to go around a corner is a direct consequence of the topological nature of the kink states.

Deviation from perfect transmission starts to occur as the magnetic field $B$ is lowered to $B <$ 6 T. Figure 2F plots the transmission coefficient of the intersection region $\tau_i$ ($B$) determined using the two-resistor model shown as the inset (see Fig. S5 for details). $\tau_i$ increases from 0.63 at $B = 0$ to unity at $B \sim 6$ T. $\tau_i$ is smaller than $\tau_{kink}$ of individual channels shown in Fig. 1H. This is not surprising because the confining bulk gaps in the intersection region are smaller and consequently non-chiral states may be present at lower energies to cause backscattering of the kink states (*11*). Increasing $\Delta$ −−our device doesn't allow this owing to gate leakage−− should enable further increase of $\tau_i$ towards unity. We discuss the current devices and improvements that can enable fully ballistic guiding of the kink states at $B = 0$ in Section 2 of (*27*).

A more powerful operation of the valley router enables it to function as a valley valve and a coherent electron beam splitter simultaneously. In this operation, the polarity of the electric field changes sign between adjacent quadrants. Kink states in opposing channels have opposite helicities, that is, states with the same chirality carry opposite valley indices K and K' and vice versa. This situation is depicted in the middle panel of Fig. 3A and leads to the suppression of straight current transmission in the absence of inter-valley scattering. This valley valve effect occurs regardless of the presence of a magnetic field and directly confirms the valley-momentum locking of the kink states. The application of a magnetic field, however, offers additional control of the wave function of the kink states. Calculations have shown that although the wave functions of the K and K′ valley kink states overlap at the CNP, where $E_F$ =0, they gradually shift in opposite directions as $E_F$ moves into the electron or hole regime (*11, 20, 28*). The shifts are illustrated in the five panels of Fig. 3A. As the kink states shift away from the mid lines of the channels, the wave function of a state coming from a particular channel has unequal overlap with states of adjacent channels, thus leading to unequal current partition. Simulations show that the wave function separation is tunable as a function of $E_F$ and $B$ and can become comparable to or greater than the width of the wave functions themselves at several Teslas (*11, 28*). Consequently, a current partition from 0 to 1 is possible.

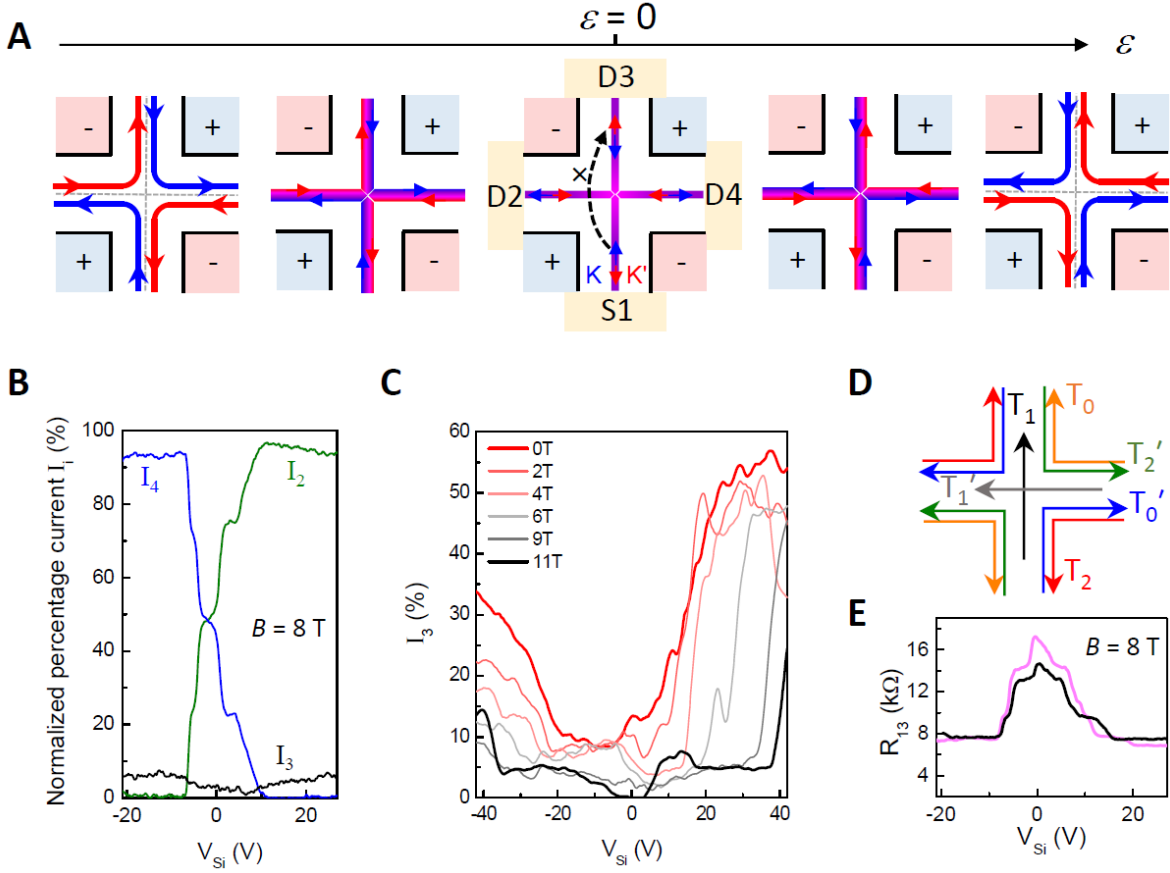

**Fig. 3. A valley valve and electron beam splitter.** (**A**) Evolution of the K and K′ kink state wave function center as a function of $E_F$ in a moderate magnetic field. The arrows mark the chirality of the kink states, from the perspective of an electron. The middle panel represents the CNP, where K and K′ states overlap. It also represents the $B = 0$ situation for all $E_F$. The dashed arrow illustrates the valley valve blocking effect. (**B**) Measurements of the normalized percentage current $I_i$ = (current to drain i/ total current) ×100% received at terminals 2 - 4 respectively as labeled in the graph while using terminal S1 as the current source. $B = 8$ T. Note the current flow is opposite of the arrows in (**A**). (**C**), $I_3$ ($V_{Si}$) at selected $B$-fields from 0 to 11 T demonstrating the robustness of the valley valve effect. (**D**) The six independent current transmission coefficients used in our model (*27*), which reflects an empirical $C_2$ rotational symmetry of our device. (**E**) Measured (magenta) and calculated (black) two-terminal resistance $R_{13}$. A contact resistance $R_c = 1$ kΩ is added to the calculated curve. $B = 8$ T. The discrepancy between theory and experiment may be caused by local microscopic imperfections of the device beyond the S-matrix model.

To test these predictions experimentally, we source a constant current from one terminal and measure the normalized percentage current $I_i$ = (current to drain i/ total current) ×100% received at the other three terminals simultaneously. Figure 3B plots $I_2, I_3, I_4$ at $B$ = 8 T using S1 labeled in Fig. 3A as the current source. Remarkably, $I_3$ remains low in the entire range of $V_{Si}$ when all four channels are in the kink regime. Similar behavior is observed in measurements using other source terminals (Fig. S7) and in device 2 (Fig. S8). The suppression of current flow between opposing terminals provides compelling evidence of the valley valve effect, which also confirms the valley-momentum locking of the kink states. As Fig. 3C shows, the valley valve effect is already very strong at $B$ = 0 with a small $I_3$ of 8% near the CNP. $I_3$ further decreases to less than 1% at $B$ = 11 T, as the magnetic field works to suppress residual inter-valley scattering in the intersection region. The transmission on/ off ratio of $I_3$ between the "through" configuration (Fig. 2A) and the blocking configuration (Fig. 3A) is approximately 8 at $B$ = 0 and more than 100 in a magnetic field. The performance of the valley valve is similar to a recently reported state-of-the-art all-electric spin valve (*29*). Unlike a spin valve, however, here the source current is nearly valley-unpolarized. The omission of the valley injection step is an advantage of the underlying topological valleytronic concept.

Instead of propagating forward, the kink state wave functions from S1 split at the intersection and propagate towards terminals D2 and D4. Both $I_2$ and $I_4$ vary co-linearly with $V_{Si}$, forming a prominent X-like feature at 8T (Fig. 3B). The tunable range of $I_2$ ($I_4$) increases with $B$ until it saturates close to the full range of 0 to 1 at ~ 5 T (Fig. S7, A and B). This current partition behavior is reproducible using different source terminals (Fig. S7C) and in different devices (Fig. S8). It is in excellent agreement with the wave function separation scenarios depicted in Figs. 3A and represents an electron analog of an optical beam splitter and a quantum point contact for the kink states. Furthermore, in Fig. S7D we show the large impact of unequal gap size on the current partition. By adjusting the size of the gap on the four quadrants at a fixed $E_F$, $I_2$ ($I_4$) can change by 50%. These results are promising steps towards the implementation of a zero-magnetic-field beam splitter (*28*).

We developed an S-matrix model (Section 8 of (*27*)) to describe the transmissions of the kink states between different channels with six independent coefficients schematically shown in Fig. 3D. These coefficients are obtained directly from measurements shown in Fig. 3B and in other source-drain setups (Fig. S7). Using the experimental input and the Landauer-Büttiker formula, we have calculated the resistance for various two-terminal and non-local measurement geometries and compared to measurements. The agreement between theory and experiment is excellent and affirms the one-dimensional transport nature of the kink states. As an example, Fig. 3E shows the calculated and measured two-terminal resistance $R_{13}$; other scenarios are discussed in Fig. S10. More fundamental understandings and predictions of the transmission process would require details of the electrostatics (*28*) and also possibly band structure effects such as trigonal wrapping (*30, 31*).

When the magnetic field increases to above 6 T, plateaus at ¼, ½, and ¾ start to appear in the transmission coefficient $T_i$. Correspondingly, conductance plateaus in integer and half-integer units of $e^2/h$ appear in the measured conductance between terminals 1 and 3 $G_{13}$. The quantization becomes increasingly prominent and precise as $B$ increases, with the data of $T_i$ and $G_{13}$ at $B = 16$ T given in Figs. 4, A and B. (See Fig. S11 for data up to 18 T). Their appearance is caused by the energy splitting of the four kink state modes in a strong magnetic field (Fig. 4B, inset) (*32*) and consequently the spatial separation of the modes in a guiding center description. The large spatial separation between the modes leads to values of $T_i$ taking either 0 or 1 for each mode, with the choice given by the position of $E_F$ with respect to the crossing point of that mode. The average of all four modes then produces quantized coefficients at multiples of ¼. This leads to conductance plateaus quantized in integer and half integer units of $e^2/h$, as Fig. 4B shows. In Section 9 of (*27*), we provide a detailed understanding of the fractional quantization of $T_i$ and its manifestation in transport, which reflect the coexisting helical and chiral nature of the kink states in a magnetic field.

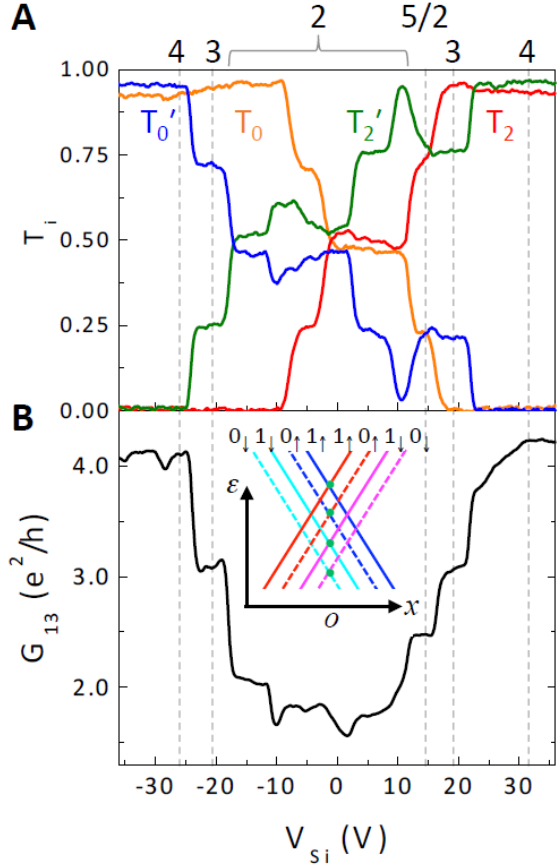

**Fig. 4. Quantized transmission coefficients in a strong magnetic field.** (**A**) Measured transmission coefficients $T_0$, $T_2$, $T_0'$ and $T_2'$ and (**B**) $G_{13}$ as a function of $V_{Si}$ at $B = 16$ T. The locations of the quantized $G_{13}$ are marked in the plot; their conductance values are labeled at the top axis in unit of $e^2/h$. Spikes in the middle of a plateau are likely caused by microscopic potential irregularities. A contact resistance of $R_c = 1174$ Ω is subtracted. The inset of (**B**) shows a guiding center description of the energy spectrum of the valley kink states in strong magnetic and electric fields.

**Acknowledgments:** We are grateful for helpful discussions with Jeffrey Teo. We thank Jan Jaroszynski of the NHMFL for experimental assistance.

**Funding:** J. L., Z. Y., and J. Z. are supported by the NSF (Grant Number DMR-1506212 and DMR-1708972). C. X. L., J. X. Z. and R.-X. Z. acknowledge the support from the Office of Naval Research (Grant Number N00014-15-1-2675). K.W. and T.T. acknowledge support from the Elemental Strategy Initiative conducted by the MEXT, Japan and the CREST (JPMJCR15F3), JST. Part of this work was performed at the NHMFL, which was supported by the NSF through NSF-DMR-1157490 and the State of Florida. Part of this work was carried out in the Nanofabrication Laboratory at Penn State's Materials Research Institute. **Author contributions:** J. Z. and J. L. conceived the experiment. J. L. designed and fabricated the devices and made the measurements. Z. Y. assisted in device fabrications. J. L. and J. Z. analyzed the data. R.-X. Z. and J. X. Z. performed the theoretical calculations. R.-X. Z. and C. X. L. analyzed theoretical results. K. W. and T. T. synthesized the *h*-BN crystals. J. Z., J. L., R.-X. Z., and C. X. L. wrote the manuscript with input from all authors. **Competing financial interests:** The authors declare no competing financial interests.

**Data and materials availability:**

Data shown in this paper are available at (*33*)

**Supplementary Materials:**

Supplementary Text

Figures S1-S14

Tables S1

Supplementary References (*34-36*)

# Supplementary Materials for

## A valley valve and electron beam splitter

Jing Li[1], Rui-Xing Zhang[1], Zhenxi Yin[1], Jianxiao Zhang[1], Kenji Watanabe[2], Takashi Taniguchi[2], Chaoxing Liu[1], Jun Zhu[1,3]

Correspondence to: jzhu@phys.psu.edu (J. Zhu)

**This PDF file includes:**

Supplementary Text
    1. Device fabrication
    2. Device characterization
    3. Comparison of the even and odd gating configurations
    4. Determining and controlling the contact resistance $R_c$
    5. The determination of $R_{kink}$, $\tau_{kink}$ and $\tau_i$.
    6. Additional properties of the electron beam splitter
    7. The valley valve and beam splitting effects in device 2
    8. The S-matrix model and comparison between experiment and theory
    9. Quantized transport in high magnetic field
Figs. S1 to S14
Tables S1

# 1. Device fabrication

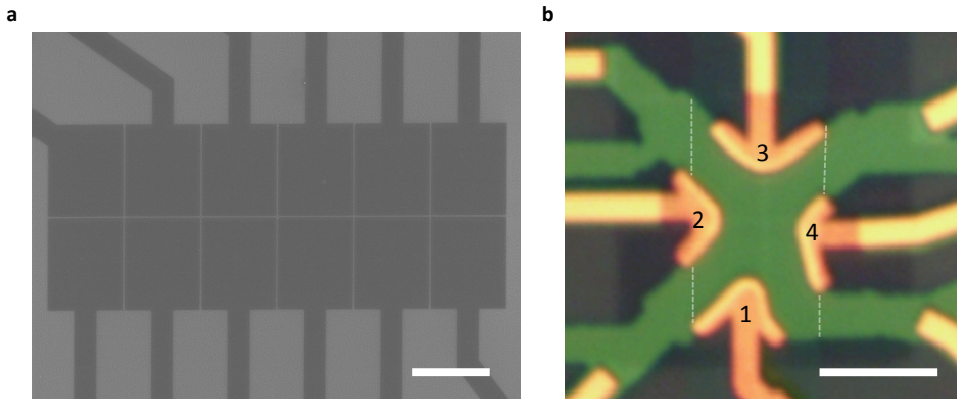

Figure S1: **a** A scanning electron microscopy (SEM) image of a bottom graphite split gate structure. **b** An optical micrograph of device 1 before the top gates are made. The dashed lines outline the edges of the bilayer graphene sheet. Scale bar is 5 μm in both images.

We first exfoliate multi-layer graphite (Kish graphite) sheets to heavily doped Si/SiO$_2$ substrate then use standard e-beam lithography and oxygen plasma etching to pattern a series of bottom split gates as shown in Fig. S1a. The width of the splits is 70 nm in both directions. We then use a van der Waals dry transfer technique (*23*) to make a *h*-BN/bilayer graphene/*h*-BN stack. The stack is transferred to the bottom split gate structure. Our experiences indicate that a AB / BA domain boundary reported by Ju et al (*17*) is very rare in our exfoliated bilayer graphene flakes. Their unique response to the gating polarity enables us to confirm their absence in our devices. We ensure the crystallographic edges of the bilayer sheet is not aligned with the direction of either split to ensure no channel is along an armchair direction. Then, the sample is annealed in an Ar/ H$_2$ (90%/10%) atmosphere at 450 °C for 3 hours to reduce bubbles introduced by the transfer process. We then examine the covered intersections in an atomic force microscope (AFM) to identify clean and bubble-free candidate for device fabrication. Following Ref. (*34*), we use a negative-tone resist hydrogen silsesquioxane (HSQ) as the etching mask, and use reactive ion etching (CHF$_3$/O$_2$ at 10:1 ratio) to define the device area shown in Fig. S1b. We then pattern and deposit the Cr/Au side contacts and electrodes that will connect to the top gates. Lastly, we pattern and deposit the four Ti/Au top gates. Using an alignment procedure detailed in Ref. (*11*), we can match the width and position of the top and bottom splits to a few nm in precision. An optical image of device 1 is shown in Fig. 1D of the text.

## 2. Device Characterization

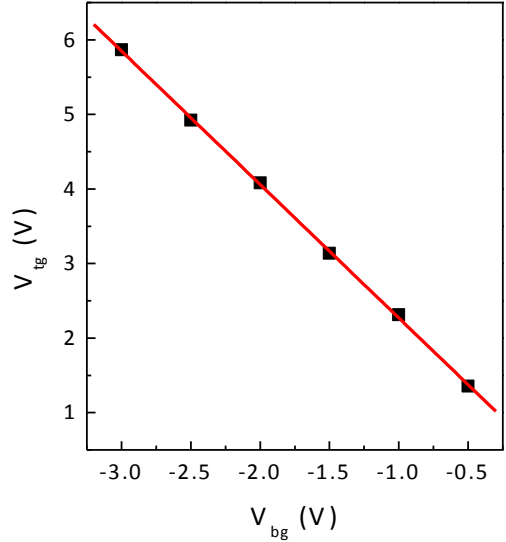

Figure S2: $V_{tg}$ - $V_{bg}$ relation for the charge neutrality point of the northwest quadrant. Red line is a linear fit with a slope of -1.8.

We employ the following steps to determine the gating efficiencies of the top and bottom gates and the gate voltages that correspond to the charge neutrality points (CNPs) of the dual-gated area in each quadrant. We first obtain an approximate $V_{tg}$ - $V_{bg}$ relation for the CNPs of all four quadrants by shorting all top gates together and all bottom gates together and perform two-terminal resistance measurements as a function of sweeping $V_{tg}$ at fixed $V_{bg}$'s, similar to measurements shown in Fig. S5c in our previous work (*11*). The resistance maxima are identified as the CNPs. We then repeat the same measurement for each quadrant by tuning its top and bottom gates (e.g. measure $R_{23}$ while changing $V_{NWt}$ and $V_{NWb}$) while setting the other three quadrants at their CNPs so that they are insulating. This procedure allows us to find the precise gate voltages of the CNPs for each quadrant. They differ slightly from each other due to slightly different environmental doping in different parts of the device while the gating efficiency ratio is found to be 1:1.8 for all our top and bottom gates. The voltage on the doped Si gate is fixed at $V_{Si}$ = 50 V in the measurements above so that the channels are conducting in parallel but do not affect the determination of the gate voltages of the CNPs. Fig. S2 plots the $V_{tg}$ - $V_{bg}$ relation for the CNP of the northwest quadrant as an example.

To determine the gating efficiencies, we measure the magneto-resistance of the northwest quadrant $R_{23}$ as a function of $V_{NWt}$ while fixing the magnetic field $B = 2$ T and $V_{NWb} = 1$ V. The other three quadrants are set at their CNPs so that they are insulating. We calculate the gating efficiency from the quantum oscillations and obtain $6.13 \times 10^{11}$ cm$^{-2}$V$^{-1}$ for the top gates in device 1. This value is in good agreement with the measured top $h$-BN thickness of 27nm and the dielectric constant $\varepsilon = 3$ for our $h$-BN. The gate efficiency of the bottom gates is calculated to be $1.1 \times 10^{12}$ cm$^{-2}$V$^{-1}$. In our measurements, gating configurations are labeled by voltages applied to the bottom gates while voltages on the top gates are set accordingly to keep the dual-gated areas at their respective CNPs. In addition, we set the magnitudes of the bottom gate voltages to be the same in all four quadrants, e.g. ± 3V. We estimate that this voltage corresponds to a displacement field of $D \sim 0.66$ V/nm and a bulk gap of $\Delta \sim 86$ meV from prior knowledge (*22*). One of the gates breaks down shortly beyond 3V. Comparing to the typical breakdown field of $D \sim 1.5$ V/nm in our dual BN-gated devices (*22*), we suspect that over etching the bottom BN flake in the side-contact making process (*23*) has reduced its breakdown field. Avoiding this in future devices should enable us to reach $\Delta \sim 200$ meV in the bulk and greatly improve the quantization of each channel, the transmission coefficient of the valve "on" state, as well as the on/off ratio of the valley valve at $B = 0$. These phenomena are discussed in detail in the text and later sections of the supplementary material.

## 3. Comparison of the even and odd gating configurations

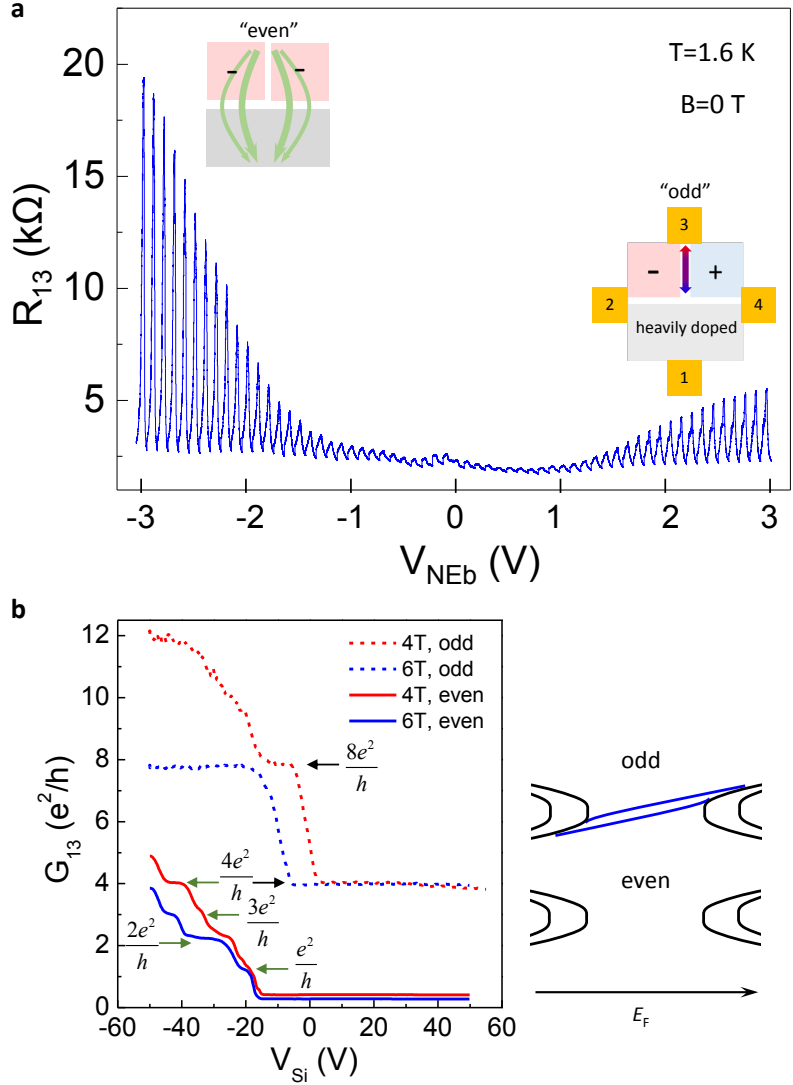

Figure S3. (a) Conductance of the north channel in the (- +) and ( - -) gating configurations. The NW quadrant is placed at the CNP with a negative gap. $V_{NWb} = -2.8$ V. $V_{NWt} = +2.8$ V. $V_{NEb}$ varies from -3 V to +3 V in 0.1 V steps. We update $V_{NEb}$ and $V_{NEt}$ simultaneously to keep the NE quadrant at the CNP while the gap sign changes from negative to positive as the diagrams show. At each set of gate voltages, we sweep $V_{si}$ from -50 V to 60V. The graph plots the 61 $V_{si}$ sweeps together. The peaks correspond to the CNP of the channel. On the left, $R_{13}$ continues to increase with increasing $\Delta$ of the NE quadrant. The finite $R_{13}$ comes from parallel conduction through the gapped bulk. The resistance is not too high because of a large width/length ratio of the bulk area (~ 14 to 1). ON the right, the presence of the kink states leads to the saturation of $R_{13}$. As an estimate, we note that a parallel resistor model of 19.3 k$\Omega$ // $h/4e^2$ and a contact resistance of $R_c = 800$ $\Omega$ yields the observed $R_{13} = 5500$ $\Omega$ at the CNP of the rightmost curve. From device 1. (b) $G_{13}$ vs $V_{si}$ in both even and odd gating configurations as labeled. From the south channel of device 2. $|V_{SEb}| = |V_{SWb}| = 3.5$ V. The application of a magnetic field leads to Landau levels and quantized conductance in the conduction and valence bands of the channel, as marked in the plot and schematically shown to the right of the graph. The conductance drops to zero in the even gating configuration and is quantized to $4e^2/h$ in the odd gating configuration when a contact resistance of $R_c = 300$ $\Omega$ is subtracted.

## 4. Determining and controlling the contact resistance $R_c$

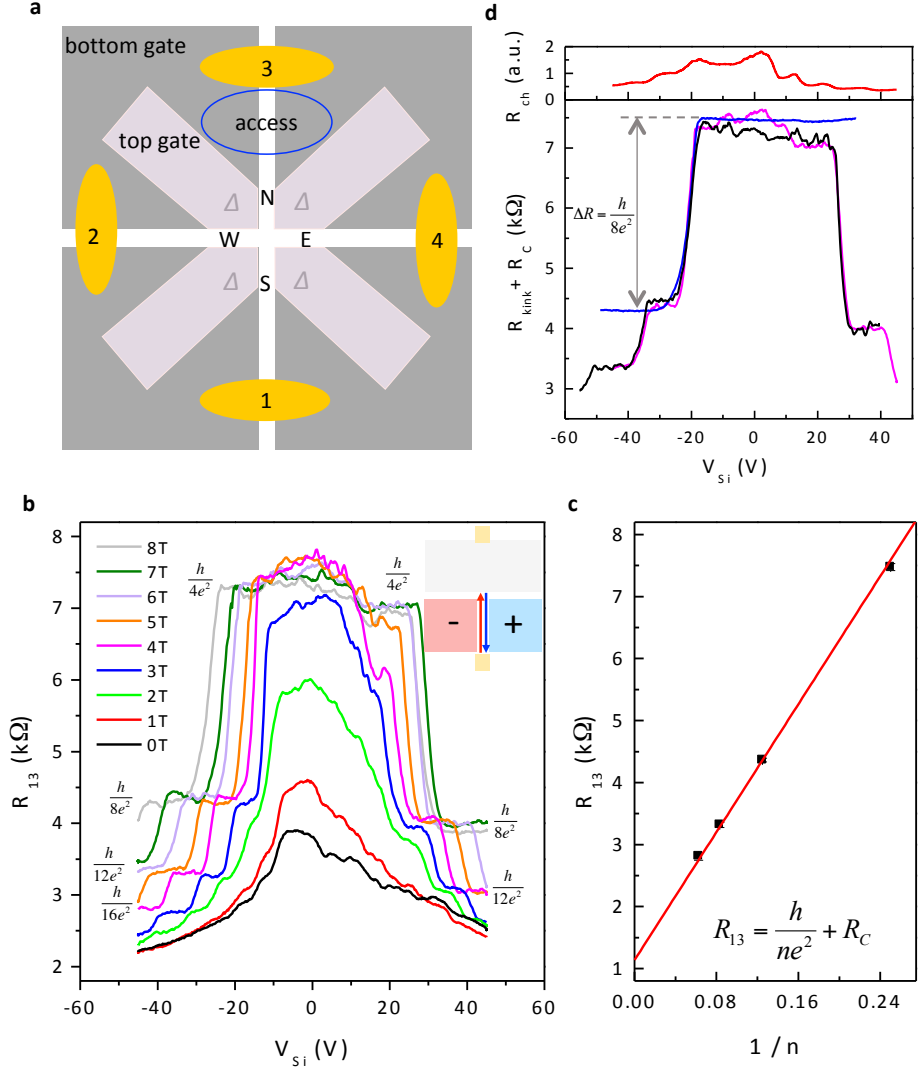

Figure S4: **a** A schematic of the contacting scheme. **b** $R_{13}$ ($V_{Si}$) at selected $B$-fields from 0 T to 8 T measuring the resistance of the south channel. The measurement scheme is shown in the inset, where the north quadrants are heavily doped and the south quadrants are gapped with the bottom gate voltage ($V_{SWb}$, $V_{SEb}$) set at (-2 V, +2 V). **c** $R_{13}$ vs the plateau index $1/n$. Points are taken from the $B = 4$ T trace on the hole side. A linear fit (red line) to the data yields an intercept $R_c = 1140$ Ω. **d** Bottom: Resistance of the kink states $R_{kink} + R_c$ for the south (magenta) and east (black) channels in device 1 and the south channel (blue) in device 2. Black and blue curves are shifted in $x$ and $y$ to overlap with the magenta curve. $B = 6$ T. Top: $R_{13}$ measured with the top and bottom gates set to 0 V. The change reflects the $V_{Si}$-dependence of the bilayer graphene channel, which includes the access region. Its profile strongly resembles that of the magenta curve in the kink regime. This led us to conclude that the $V_{Si}$ – dependence of the magenta curve in this regime is due to $R_c$.

Figure S4a shows a schematic of how the kink states are contacted. Each channel is connected to the Cr/ Au side contact through an access region, which is gated by two bottom gates and the Si backgate. The contact resistance $R_c$ combines the resistance of the metal/graphene interface and the resistance of the bilayer graphene access region. In measurements described in Fig. 2 of the text, we keep the same gate voltages on the bottom gates when measuring the north channel, the south channel and the "through" configuration and use the top gates to control doping. This strategy ensures the contact resistance included in $R_{13}$ is roughly the same for $R_N$, $R_S$ and $R_{through}$. This is indeed the case as the data in Fig. 2D shows.

Figure S4b shows measurements of the south channel $R_s$ at magnetic fields $B = 0$ to 8T. The resistance plateaus correspond to an integer number of edge states (kink + quantum Hall) contributing to the conduction, as labeled in Fig. S4b. Away from $V_{Si} = 0$ V, their resistances are well described by $R = h/ne^2 + R_c$, where the index n is the number of edge sates. Fig. S4c plots the resistance of the four plateaus from the hole side of the $B = 4$ T curve, as a function of $1/n$. A linear fit through the data indicates $R_c$ is roughly a constant for all four plateaus and $R_c = 1140$ Ω. This corresponds to ~ 600 Ω per contact. We attribute the low contact resistance to the large width of the metal/graphene interface and the heavy doping of the access region. We also see that the well-developed plateaus remain at the same value as B increases. This shows that $R_c$ is field independent.

Because a portion of the access region (the 70 nm-wide channel) is controlled by the Si backgate, $R_c$ can exhibit a noticeable dependence on $V_{Si}$ and this is included in the measurement of the channel resistance. Away from $V_{Si} = 0$, this dependence manifests as an electron-hole asymmetry, as shown by the systematically lower $R_{13}$ values in Fig. S4b on the electron side. Near $V_{Si} = 0$, $R_c$ may increase by a few hundred Ω. This also leads to an increase in the measured $R_{13}$, as shown in Fig. S4b. To examine the impact of $R_c$, Fig. S4d plots the measured resistance of several kink channels from two devices together. $R_c$ is nearly a constant in the entire range of $V_{Si}$ for the blue curve (south channel of device 2). This leads to well quantized resistance plateaus for the kink states and the quantum Hall states, with the difference between the two very close to $h/4e^2$. The magenta curve (south channel in device 1), on the other hand, exhibits a maximum near $V_{Si} = 0$ V. By measuring the resistance of the bilayer graphene channel from contacts 3 to 1 independently and comparing its shape to that of the magenta curve, we conclude that the south channel most likely remains quantized in the entire kink regime while $R_c$ ($V_{Si}$) accounts for the excess over $h/4e^2$. As a third example, we show that the black curve (east channel in device 1) exhibits a slowly decreasing $R_c$, resulting in a tilted plateau in the kink regime.

## 5. The determination of $R_{kink}$, $\tau_{kink}$ and $\tau_i$.

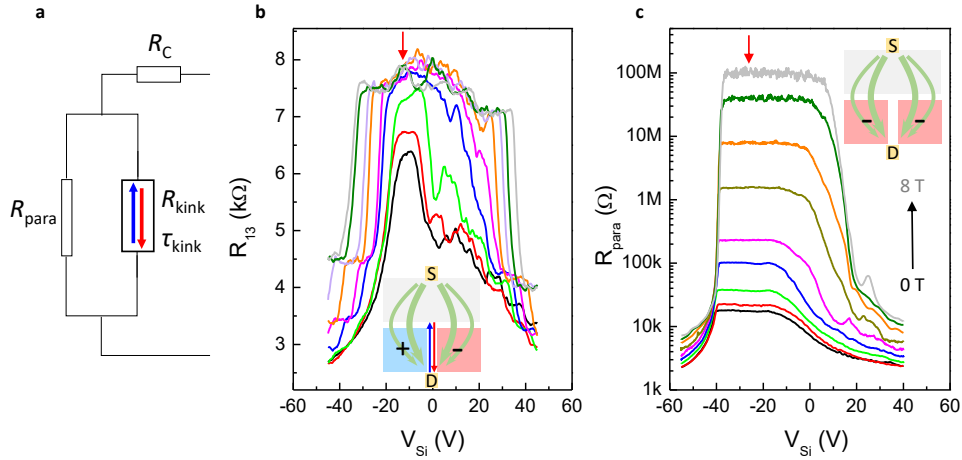

Figure S5: **a** A schematic of the two-channel model. **b** and **c**, $R_{13}$ ($V_{Si}$) at selected $B$-fields from 0 T to 8 T. The color scheme follows that of Fig. S4b. Insets show the measurement schemes and illustrate the parallel conduction paths through the gapped quadrants. $R_{para}$ remains the same in **b** and **c** while the kink states are absent in **c**. We use $R_{para}$ measured in **c**, $R_{13}$ measured in **b**, together with $R_c = (1440 \pm 70)$ Ω to calculate $R_{kink}$ of the south channel at $V_{Si} = -11$ V (red arrow). ($V_{SWb}$, $V_{SEb}$) = (+3 V, -3 V) in **b** and (-3 V, -3 V) in **c**. From device 1.

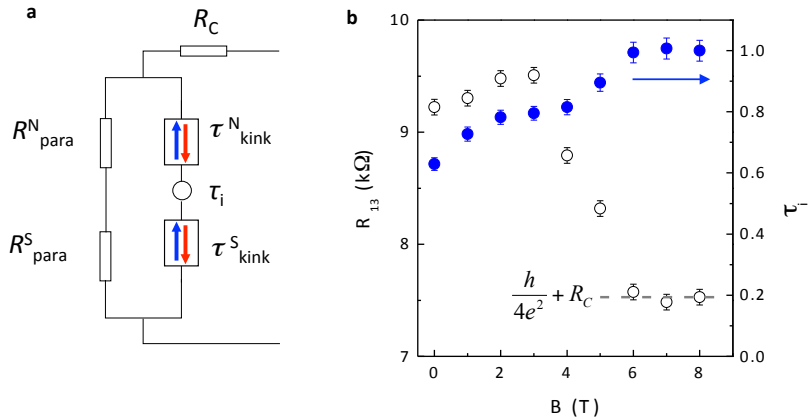

Figure S6: **a** A schematic of the model used to determine the transmission coefficient $\tau_i$ of the intersection region. **b** The measured $R_{13}$ in the "through" configuration and the calculated $\tau_i$ as a function of the magnetic field. $R_c = 1075$ Ω. ($V_{SWb}$, $V_{SEb}$) = (-3 V, +3 V). ($V_{NWb}$, $V_{NEb}$) = (-3 V, +3 V). From device 1.

In data shown in Fig. 1G of the text, the kink state resistance drops below $h/4e^2$ at $B = 0$ and 2T. This is due to finite parallel conduction. We use a two-resistor model first established in Ref. (*11*) (Fig. S19) and illustrated in Fig. S5a to calculate the kink state resistance $R_{kink}$, and the transmission coefficient $\tau_{kink} = \frac{h/4e^2}{R_{kink}}$. $R_{papa}$ represents the resistance of the parallel conduction paths. $R_{para}$ can be simulated by an "even" field configuration as shown in the inset of Fig. S5c. Using data shown in Figs. S5b and 5c, we can extract the resistance of the kink states at $V_{Si} = -11$ V (red arrow in 5b) using $R_{kink} = \frac{(R_{13}-R_C)R_{para}}{R_{para}-(R_{13}-R_C)}$. Here $R_c = R_{13} - h/4e^2 = (1440 \pm 70)$ Ω is the contact resistance. Similar analyses are done for two other channels and the results are plotted in Fig. 1H of the text.

We use a modified two-resistor model shown in Fig. S6a to extract the transmission coefficient of the intersection $\tau_i$ in the "through" configuration. $R_{13} = R_C + \left(\sum R_{para} // \frac{h/4e^2}{\tau^N_{kink}\tau_i\tau^N_{kink}}\right)$, where $R^N_{para}$, $R^S_{para}$, $\tau^N_{kink}$ and $\tau^S_{kink}$ are obtained from measurements described above. Fig. S6b plots the measured $R_{13}$ and the corresponding $\tau_i$. $\tau_i$ increases from 0.63 at $B = 0$ to 1 at $B = 6$ T. The results of $\tau_i$ are also shown in Fig. 2F of the text. As discussed in length in our earlier work (*11*), the presence of charge puddles could mediate valley mixing of the kink states. In the "on" state of the valve, the confining gap in the center of the cross is reduced to about ¼ of the value in the bulk. The reduction of $\tau_i$ is likely due to enhanced coupling to the charge puddles there. Further increase of the bulk gap, as discussed in Section 2, should eliminate this mechanism to enable perfect transmission at $B = 0$.

## 6. Additional properties of the electron beam splitter

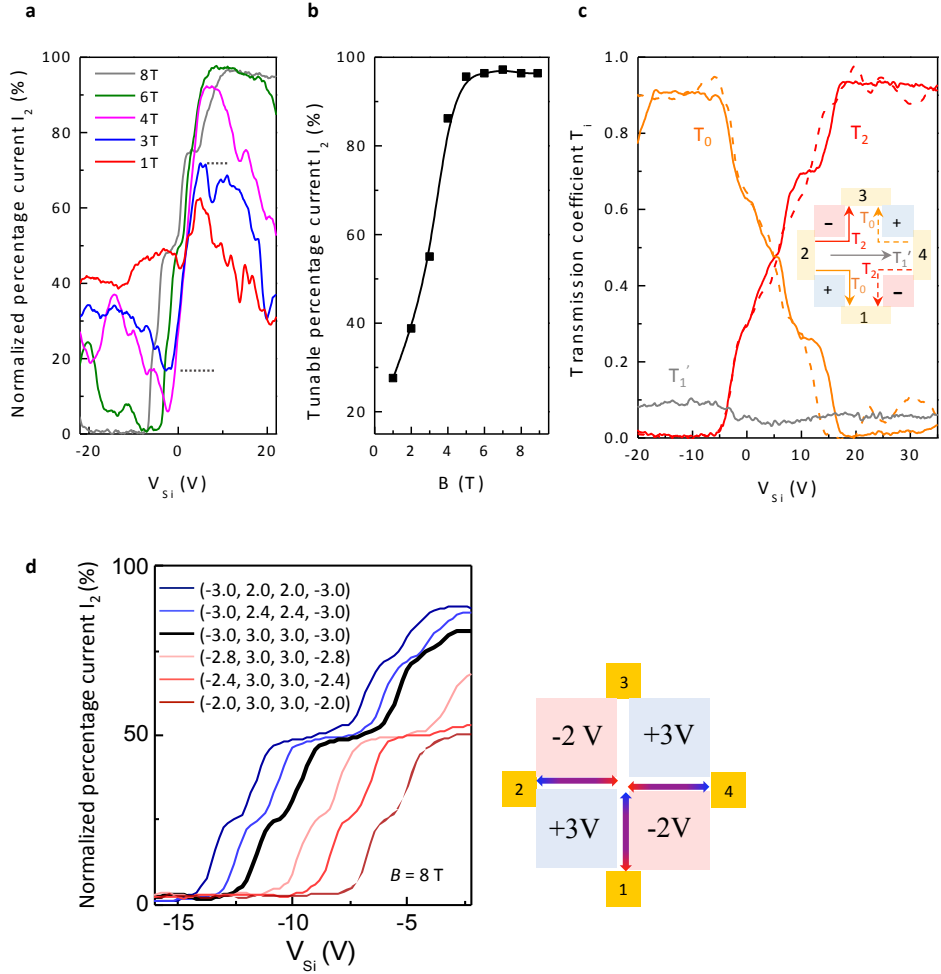

Figure S7: Properties of the beam splitter. **a** Normalized percentage current $I_2$ at selected $B$ - fields from 1 to 8 T. Dashed lines mark the tunable range of $I_2$ for the 3 T data. **b** The tunable range of $I_2$ as a function of the magnetic field. **c** Transmission coefficients $T_0$, $T_2$ and $T_1^{'}$ measured using terminal 2 (solid curves) or 4 (dashed curves) as the current source. The color scheme follows that of the inset. The excellent agreement between the solid and dashed curves indicates an approximate $C_2$ rotational symmetry. ($V_{NWb}$, $V_{NEb}$, $V_{SWb}$, $V_{SEb}$) = (-3 V, +3 V, +3 V, -3 V). $B$ = 8 T. **d** The combined effect of $\Delta$ and $E_F$ on current partition. Normalized percentage current $I_2$ at fixed ($V_{NWb}$, $V_{NEb}$, $V_{SWb}$, $V_{SEb}$) values as labeled in the plot. Unequal gap sizes shift the kink states towards the quadrants with smaller gaps. The diagram to the right illustrates the situation corresponding to the rightmost trace. Here, transmission towards terminal 2 is suppressed, in agreement with the data.

In this section, we provide more data and analysis of the beam splitting operation shown in Fig. 3 of the text. Expanding the data shown in Fig. 3B of the text, Fig. S7a plots the normalized percentage current $I_2$ at selected magnetic fields from 1 T to 8 T. $I_2$ is linear with $V_{Si}$ near the equipartition point with a positive slope. Both the slope and the tunable range expanded by $I_2$ increase with increasing $B$ at small $B$ but quickly saturate. This behavior is in good agreement with the numerical calculations of Wang et al (*27*). We note that the trending of $I_2$ towards 50% after peaking is due to the presence of conduction band quantum Hall edge states in both the south and north channels, which diverts 50% of the current towards $I_3$ (see Fig. 3C of the text for the behavior of $I_3$). It is not related to the functioning of the beam splitter. Fig. S7b plots the tunable range of $I_2$ vs. $B$. It increases from 28% at $B = 1$ T to 96% at $B = 5$ T.

We determine the transmission coefficient $T_i$ from measurements conducted using different terminals as the current source. Ideally a $C_4$ rotational symmetry is expected. However our data show an approximate $C_2$ rotational symmetry, as illustrated in Fig. S7c. $T_0$ and $T_2$ shown in Fig. S7c are different from $T_0'$ and $T_2'$, which are measured while sourcing the current from terminal 1 (Fig. 3B of the text). We suspect that the $C_2$ symmetry arises from the unequal gap size $\Delta$ on the four quadrants. Our measurements in Section S2 lack the resolution to precisely locate the $\Delta = 0$ condition for each quadrant. If the $\Delta = 0$ condition corresponds to positive offsets on the bottom gate voltages (The offsets are usually within ± 0.2 V from our experiences), ($V_{NWb}$, $V_{NEb}$, $V_{SWb}$, $V_{SEb}$) = (-3 V, +3 V, +3 V, -3V) would lead to larger $\Delta$ on the NW and SE quadrants and smaller $\Delta$ on the NE and SW quadrants. The effect of unequal $\Delta$ is further explored in Fig. S7d, where we show the current partition can be continuously tuned by a large percentage by adjusting the ratio of the gaps. The underlying mechanism of this tuning comes from the shift of the zero-potential line towards the quadrants with smaller gaps and was predicted in Ref. (*27*). Fig. S7d is a promising step towards the realization of a tunable beam splitter at $B = 0$.

## 7. The valley valve and beam splitting effect in device 2

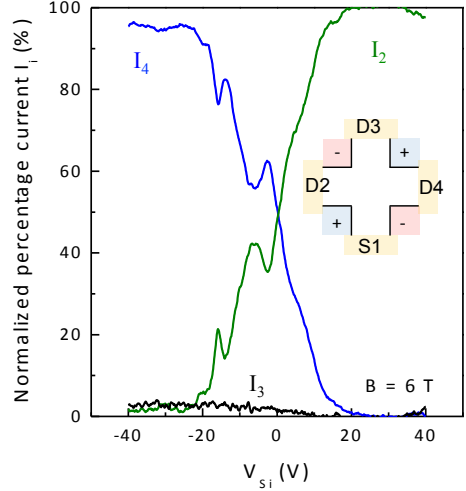

Figure S8: Normalized percentage current measured in device 2. ($V_{NWb}$, $V_{NEb}$, $V_{SWb}$, $V_{SEb}$) = (-3.5 V, +3.5 V, +3.5 V, -3.5 V). $B = 6$ T.

The valley valve and beam splitting effects observed in device 1 is reproducible in device 2. Figure S8 plots an example at $B = 6$ T. $I_3$ is nearly zero in the entire range and $I_2$ and $I_4$ exhibit current partition behavior similar to that of device 1 shown in Fig. 3 of the text.

## 8. The S-matrix model and comparison between experiment and theory

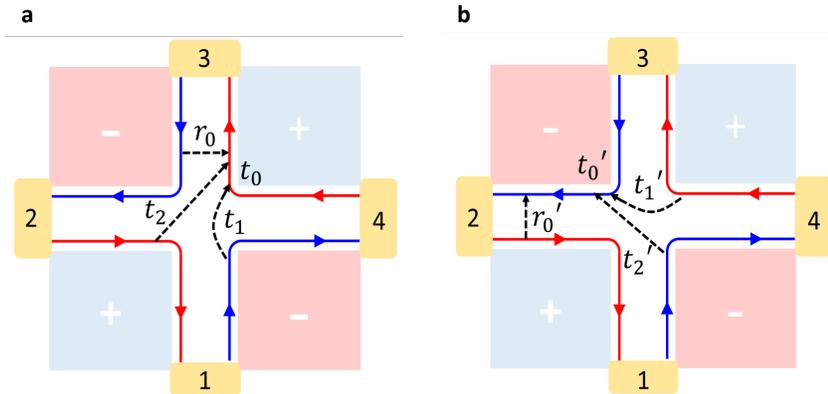

Figure S9: **a** and **b** The S-matrix model describing the transmission of kink states between different channels. Two sets of parameters are used to describe the scattering amplitude as illustrated in the diagrams. The diagrams are drawn for $\varepsilon > 0$ (See Fig. 3A of the text). Blue (red) arrows denote the kink states at K (K′).

We adopt an S-matrix model to relate the outgoing modes $\Psi_{out} = (\psi_{1,out}, \psi_{2,out}, \psi_{3,out}, \psi_{4,out})^T$ with the incoming modes $\Psi_{in} = (\psi_{1,in}, \psi_{2,in}, \psi_{3,in}, \psi_{4,in})^T$ through $\Psi_{out} = S\,\Psi_{in}$ where $\psi_{i,in/out}$ denotes the in/out mode for the $i$-th terminal and

$$S = \begin{pmatrix} r_0 & t_0 & t_1 & t_2 \\ t_2' & r_0' & t_0' & t_1' \\ t_1 & t_2 & r_0 & t_0 \\ t_0' & t_1' & t_2' & r_0' \end{pmatrix}.$$

Figures S9a and 9b illustrate the possible scattering processes between all terminals, constrained by the experimentally observed $C_2$ symmetry. The conductance matrix $G$ is related to the S-matrix by

$$G_{i \leftarrow j} = \frac{4e^2}{h}|S_{ij}|^2 \tag{S1}$$

where the coefficient $4\,e^2/h$ accounts for the four modes of the kink states in each channel (two due to spin, two due to layer isospin). Using Eq. (S1) and the Landauer-Büttiker formula shown in Eq. (S2)

$$I_i = \sum_j (V_i - V_j) G_{i \leftarrow j} \tag{S2}$$

we obtain

$$\begin{pmatrix} I_1 \\ I_2 \\ I_3 \\ I_4 \end{pmatrix} = \frac{4e^2}{h} \times \begin{pmatrix} \tau & -T_0 & -T_1 & -T_2 \\ -T_2' & \tau' & -T_0' & -T_1' \\ -T_1 & -T_2 & \tau & -T_0 \\ -T_0' & -T_1' & -T_2' & \tau' \end{pmatrix} \begin{pmatrix} V_1 \\ V_2 \\ V_3 \\ V_4 \end{pmatrix}, \tag{S3}$$

where $T_i = |t_i|^2$ and $T_i' = |t_i'|^2$ for $i \in \{0,1,2\}$, and $\tau = T_0 + T_1 + T_2 = 1 - |r_0|^2$ and $\tau' = T_0' + T_1' + T_2' = 1 - |r_0'|^2$. The current transmission coefficients $T_i$'s are shown in Fig. 3D. By setting $V_1 = V$ and $V_2 = V_3 = V_4 = 0$, we find that $\tau = \frac{h}{4e^2}\frac{I_1}{V}$, $T_0' = -\frac{I_4}{I_1}\tau$, $T_1 = -\frac{I_3}{I_1}\tau$, $T_2' = -\frac{I_2}{I_1}\tau$.

Our measurements are in the ballistic regime with $\tau \approx \tau' \approx 1$. Therefore, the transmission coefficients can be directly obtained from the measured percentage current as described in Fig. 3 of the text. Equation (S3) allows us to predict the outcomes of various measurement geometries using the experimentally obtained coefficients as input, and compare to the corresponding measurements. The results on the two-terminal resistance $R_{13}$ are given in Fig. 3E of the text. Figures S10a and b show additional examples. In all cases, theory and experiment reach excellent agreement. As Figs. S10a and b show, both experiment and theory show $R_{43} < R_{14}$, as a consequence of the $C_2$ symmetry (See Section S6 for more discussions). It is also worthwhile pointing out that $R_{43}$ and the corresponding non-local resistance $R_{12,43}$ (and similar $R_{14}$ and $R_{23,14}$) follow a simple relation in the entire kink regime, which reads

$$R_{43} - R_{12,43} = \frac{h}{4e^2} + R_c. \tag{S4}$$

This relation is successfully reproduced in our calculated curves (without $R_c$) as well. In the next section, we rewrite the Landauer-Büttiker formula in terms of quantized transmission coefficients. We can show that the two-terminal conductance $G_{43}$ is $G_{43} = \frac{8+MN-2(M+N)}{16+MN-4N} \frac{8e^2}{h}$, while its partner non-local conductance $G_{12,43}$ is given by $G_{12,43} = \left(\frac{16}{M} + \frac{16}{4-N} - 8\right) \frac{e^2}{h}$. It is then straightforward to show that Eq. (S4) is satisfied analytically.

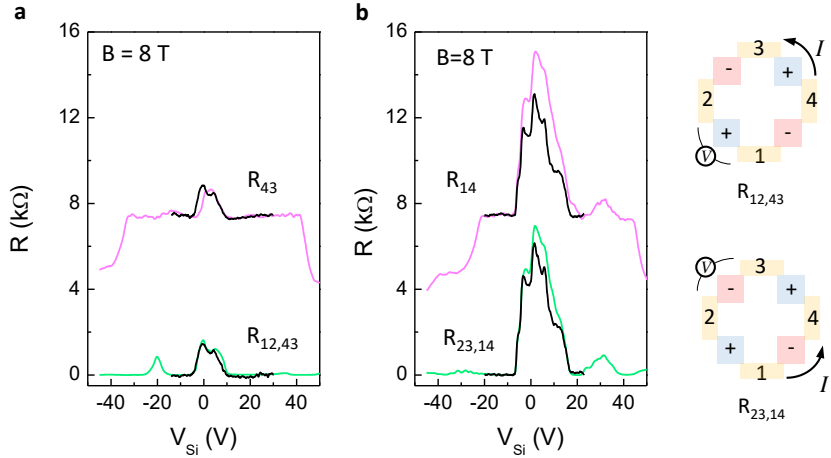

Figure S10: Comparison between theory (black curves) and experiment (magenta and green curves) on two-terminal resistances $R_{43}$, $R_{14}$ and the partner non-local resistances $R_{12,43}$ and $R_{23,14}$. In $R_{43}$ and $R_{14}$, a constant $R_c$ is added to the theoretical curves to match data. Diagrams illustrate how the non-local resistances are measured.

## 9. Quantized transport at high magnetic field

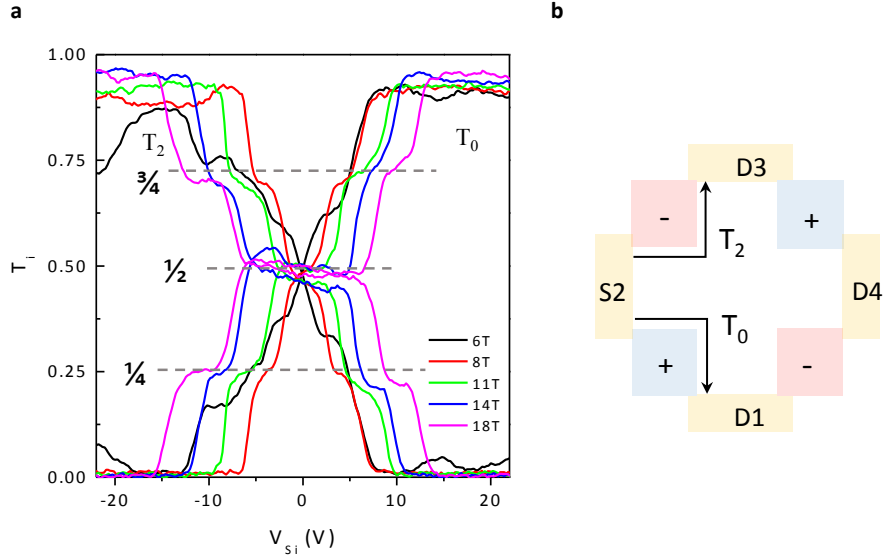

Figure S11: **a** Quantized transmission coefficients $T_0$ and $T_2$ at fractional values marked in the plot at selected magnetic fields from 6 to 18 T. The centers of the curves are aligned to facilitate comparison. **b** The measurement setup.

Figure S11 plots transmission coefficients $T_0$ and $T_2$ at fixed magnetic fields from 6 T to 18 T, where quantizations at fractional values of ¼, ½ and ¾ are clearly seen. These observations can be understood by taking into account the Landau level quantization of the bulk bands of bilayer graphene (*35*) and the resulting evolution of the kink states to valley kink states in a strong magnetic field (*32, 36*). Figure S12a shows a likely energy spectrum of the valley kink states in a guiding center description for the north channel. The energy splitting between the different modes of the valley kink states results in four distinct crossing points of the states from K and K′ valleys. As $E_F$ sweeps through the crossing points, the physical separation between the K and K′ states is different in size and direction for each mode; thus each mode can take on different values of transmission coefficients. As the crossing points grow farther apart in stronger magnetic fields, the attainment of the 0 or 1 limit for each of the four modes manifests as plateaus of fractional coefficients at 0, ¼, ½, ¾, and 1. This process is illustrated in Fig. S12 b-f.

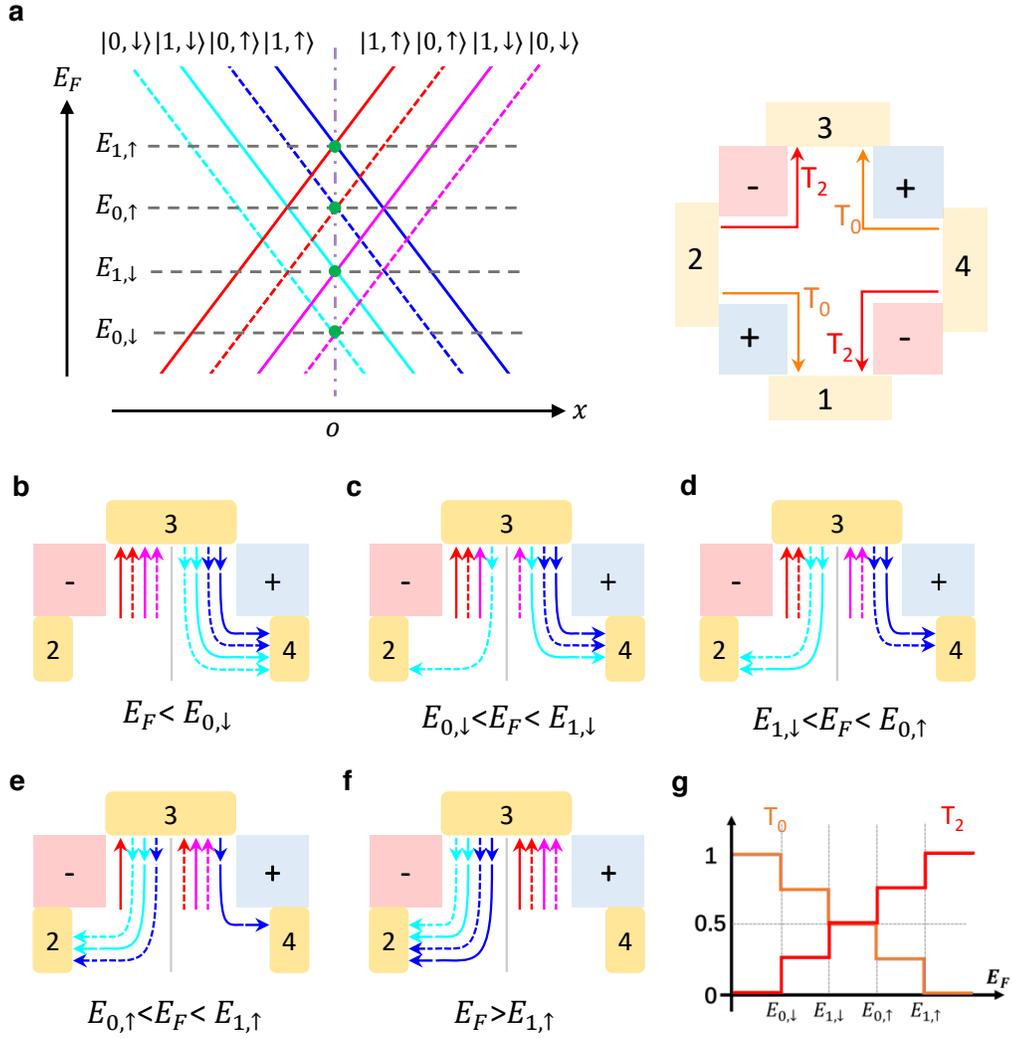

Figure S12 Fractionalized transmission coefficients. **a** An energy spectrum of the valley kink states in strong magnetic and electric fields in a guiding center description. **b to f** The evolution and quantization of $T_0$ as a function of $E_F$. $T_0$ represents current transmission from terminal 4 to 3, or equivalently K valley states flowing from terminal 3 to 4. The connections are made based on the physical distance between modes. A connecting line represents a transmission probability of 1. **g** An idealized plot of $T_0$ and $T_2$ as a function of $E_F$. Similar analysis applies to $T_0'$ and $T_2'$. Here we have neglected the effect of intervalley scattering and assumed that transmissions preserve the spin and isospin indices of the valley kink states.

We can rewrite the two-terminal conductance $G_{13} = 4e^2/h\,(1 - T_0 T_2' - T_2 T_0')$ obtained in the Landauer–Büttiker theory in terms of quantized coefficients:

$$T_0 = \frac{M}{4},\quad T_2 = \frac{4-M}{4},\quad T_0' = \frac{N}{4},\quad T_2' = \frac{4-N}{4},\quad M, N \in \{0,1,2,3,4\},$$

where $M$ is the number of modes flowing from terminal 3 (1) to terminal 4 (2) and $N$ is the number of modes flowing from terminal 2 (4) to terminal 3 (1). We obtain

$$G_{13} = \left[4 + \frac{MN}{4} - (M+N)\right]\frac{e^2}{h}. \tag{S5}$$

Table S1 lists all possible values of $G_{13}$ for given values of $T_0(M)$ and $T_0'(N)$. Equation (S5) makes it clear that the quantization of $G_{13}$ directly results from the quantized transmission coefficients. Figure S13a plots the measured coefficients and $G_{13}$ at $B = 16$ T. As Figure S13b illustrates, the agreement between experiment and Eq. (S5) is excellent.

| $G_{13}\ (\frac{e^2}{h})$ | $T_0 = 0$ ($M = 0$) | $T_0 = 1/4$ ($M = 1$) | $T_0 = 1/2$ ($M = 2$) | $T_0 = 3/4$ ($M = 3$) | $T_0 = 1$ ($M = 4$) |
|---|---|---|---|---|---|
| $T_0' = 0$ ($N = 0$) | **4** | 3 | 2 | 1 | 0 |
| $T_0' = 1/4$ ($N = 1$) | **3** | **2.5** | **2** | 1.5 | 1 |
| $T_0' = 1/2$ ($N = 2$) | 2 | 2 | **2** | **2** | **2** |
| $T_0' = 3/4$ ($N = 3$) | 1 | 1.5 | **2** | **2.5** | **3** |
| $T_0' = 1$ ($N = 4$) | 0 | 1 | 2 | 3 | **4** |

Table S1: Possible values of $G_{13}$ as a combination of $M$ and $N$.

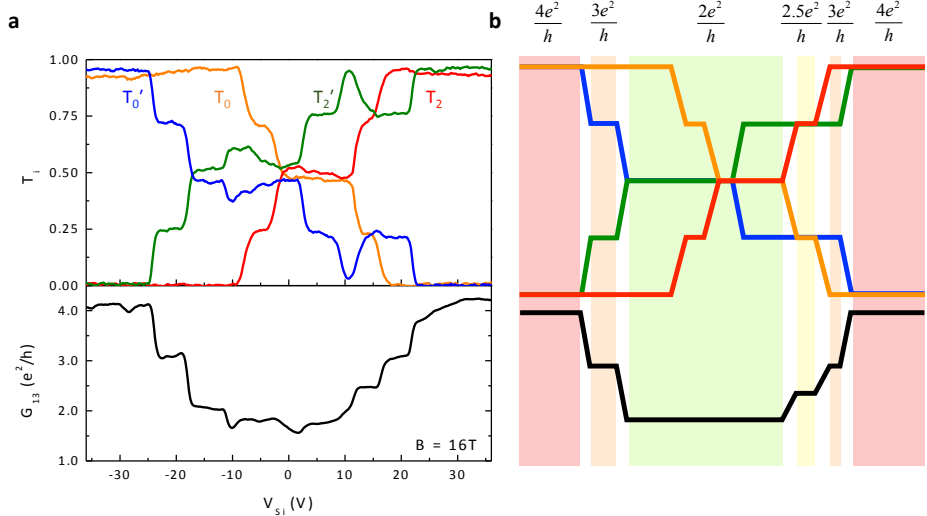

Figure S13: **a** Measured transmission coefficients $T_0$, $T_2$, $T_0'$ and $T_2'$ (top) and $G_{13}$ (bottom) as a function of $V_{Si}$ at $B = 16$ T. $G_{13} = 1/(R_{13}-R_c)$. $R_C = 1174\ \Omega$. "Glitches" in the middle of a plateau are likely due to microscopic potential irregularities. **b** Idealized data in **a** demonstrating the working of Table S1. The color scheme follows that of Table S1. Experimentally observed plateaus in $G_{13}$ are shown in bold in Table S1.

In what follows, we use the microscopic guiding center picture to provide a physical and intuitive interpretation of the quantization in $G_{13}$ ($R_{13}$). We examine two scenarios, which correspond to $M = 4$, $N = 4$ ($T_0 = T_0' = 1$) and $M = 3$, $N = 2$ ($T_0 = 3/4$, $T_0' = 1/2$) respectively, to illustrate the essential points of the picture.

Figure S14b depicts the flow of the kink states in the case of $T_0 = T_0' = 1$ and $T_2 = T_2' = 0$. Although the incoming and outgoing modes carry different valley indices, this situation resembles that of a chiral-edge-state system in transport and it is straightforward to obtain $G_{13} = 4\ e^2/h$ since four edge modes are present. $G_{13}$ of the scenario illustrated in Fig. S14c for $T_0 = 3/4$, $T_2 = 1/4$, $T_0' = 1/2$ and $T_2' = 1/2$ is less obvious. In this case, it is helpful to view it as the sum of three subsystems as shown in Fig. S14d-f. A single chiral mode contributes $e^2/h$ conductance. The resistance of the helical system is $(h/e^2 + h/e^2) // (h/e^2 + h/e^2) = h/e^2$, with each pair of counter-propagating modes contributing to a quantum resistance of $h/e^2$. The third subsystem does not contribute to $G_{13}$ since no mode connects terminals 1 and 3. Together, we recover $G_{13} = 2\ e^2/h$. The same method can be used to understand other plateaus in $G_{13}$.

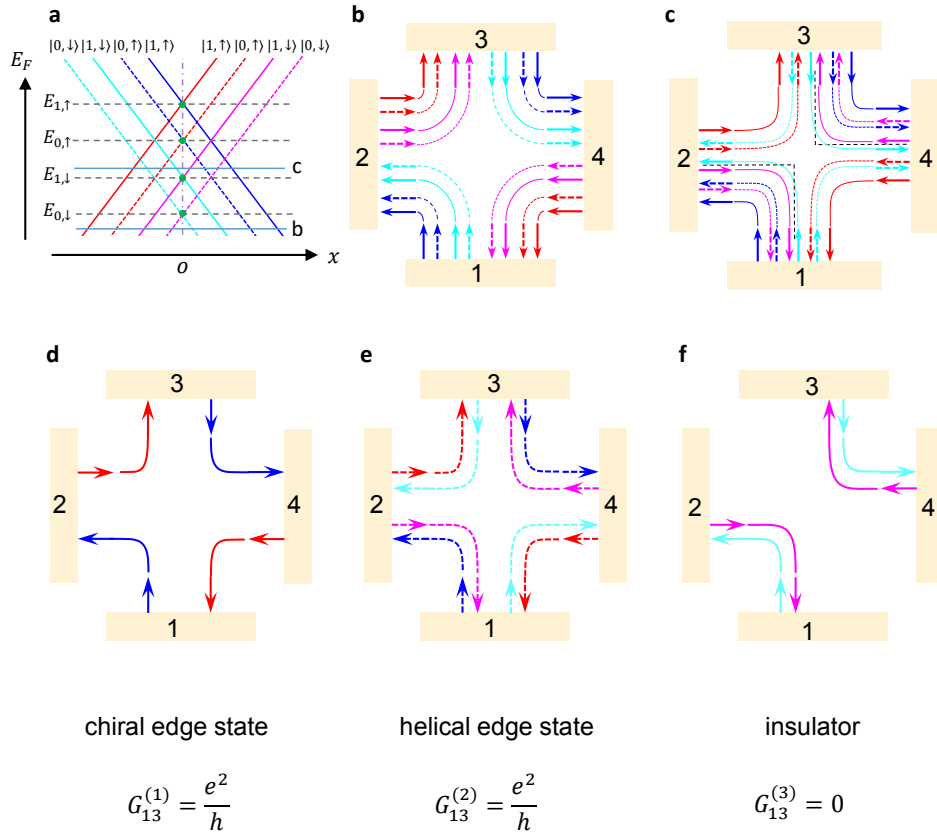

Figure S14: Quantization of $R_{13}$. **a** The energy diagram of the valley kink states shown in Fig. S11a. The blue lines mark $E_F$ in **b** and **c**. **b** The transmission scenario for $T_0 = T_0' = 1$. This situation mimics that of a chiral quantum Hall system with $G_{13} = 4\ e^2/h$. **c** $T_0 = 3/4$. $T_0' = 1/2$. The black dashed lines illustrate the shift of the crossing points towards the NE and SW quadrants, as discussed in Section 5. This shift can account for the difference between $T_0$ and $T_0'$, as the diagram illustrates. **d-f** Decomposition of **c** into the sum of three subsystems, the conductance of which add up to the observed $2\ e^2/h$.